\documentclass[sn-aps]{sn-jnl}

\usepackage{graphicx}%
\usepackage{multirow}%
\usepackage{amsmath,amssymb,amsfonts}%
\usepackage{amsthm}%
\usepackage{mathrsfs}%
\usepackage[title]{appendix}%
\usepackage{xcolor}%
\usepackage{textcomp}%
\usepackage{manyfoot}%
\usepackage{booktabs}%
\usepackage{algorithm}%
\usepackage{algorithmicx}%
\usepackage{algpseudocode}%
\usepackage{listings}%

\begin{document}

\title[Article Title]{Near-field Strong Coupling and Entanglement of Quantum Emitters for Room-temperature Quantum Technologies}

\author[1]{\fnm{Daniel D. A.} \sur{Clarke}}

\author*[1]{\fnm{Ortwin} \sur{Hess}}\email{ortwin.hess@tcd.ie}

\affil[1]{School of Physics and CRANN Institute, Trinity College Dublin, Dublin 2, D02 PN40, Ireland}


\abstract{
In recent years, quantum nanophotonics has forged a rich nexus of nanotechnology with photonic quantum information processing, offering remarkable prospects for advancing quantum technologies beyond their current technical limits in terms of physical compactness, energy efficiency, operation speed, temperature robustness and scalability. In this perspective, we highlight a number of recent studies that reveal the especially compelling potential of nanoplasmonic cavity quantum electrodynamics for driving quantum technologies down to nanoscale spatial and ultrafast temporal regimes, whilst elevating them to ambient temperatures. Our perspective encompasses innovative proposals for quantum plasmonic biosensing, driving ultrafast single-photon emission and achieving near-field multipartite entanglement in the strong coupling regime, with a notable emphasis on the use of industry-grade devices. We conclude with an outlook emphasizing how the bespoke characteristics and functionalities of plasmonic devices are shaping contemporary research directives in ultrafast and room-temperature quantum nanotechnologies.
}


\maketitle

\section{Introduction}

Coherent interactions between a single quantum emitter and an optical cavity have long resided at the heart of quantum optics \cite{mabuchi2002cqed} .
On the one hand, optical cavities provide an important platform for fundamental explorations of quantum mechanical phenomena and their philosophical implications; on the other, they are instrumentally crucial for the development of revolutionary quantum technologies that harness the interplay of light and matter, extending from photonic quantum computing and cryptography to sensing and metrology.
In particular, when the cavity-emitter energy exchange rate $g$ exceeds the intrinsic cavity loss and emitter decoherence rates, the hybrid system enters the so-called strong coupling regime. Here, the cavity photon mode and the matter electronic or phononic excitation no longer retain their separate identities, but become inextricably intertwined in the form of dressed or polariton states. The formation of these polaritons gives rise to the characteristic phenomenology of cavity quantum electrodynamics (cQED), including vacuum Rabi splitting, Rabi oscillations, modifications of the Purcell effect and non-classical photon statistics \cite{wallsmilburnbook}.
Crucially, the strong coupling regime in the single-emitter limit has emerged as a particularly promising resource for nascent photonic quantum information processing strategies, supporting quintessential functionalities like single-qubit coherent control \cite{wallraff2004strong}, ultrafast single-photon emission \cite{bello2020cavityfree} and optical switching \cite{volz2012ultrafast}, as well as quantum sensing \cite{kongsuwan2019quantum}.

Traditionally, the fragility of strongly coupled states of light and matter has demanded both high quality-factor resonators and cryogenic cooling, significantly impacting both the cost and practicality of the large-scale deployment of quantum-device technologies. The rise of quantum nanophotonics, however, has created a rich nexus of nanotechnology with photonic quantum information processing (QIP), offering prospects for advancing quantum technologies beyond their current technical limits, including their physical compactness, energy efficiency, ultrafast operability, temperature robustness and scalability.
Among photonic nanodevices, plasmonic nanoresonators are particularly special: contrary to their traditional dielectric counterparts, they offer the unique ability to confine light to extremely sub-wavelength volumes and massively enhance local electromagnetic fields via resonant surface plasmon modes, thereby constituting exceptional architectures for quantum light-matter interaction and the exploration of extreme nano-optics \cite{baranov2018novel,baumbergetal2019rev}. Crucially, room-temperature strong coupling of single molecules and colloidal quantum dots in nanoplasmonic environments has been realized using ultrathin (${\sim}\,1$ nm) metal-insulator-metal nanocavities \cite{chikkaraddy2016single,santhosh2016vacuum} and scanning probe tips \cite{gross2018near,park2019tip},
whose ultralow mode volumes of $V_{\rm m} < 100$ nm$^{3}$ enable high coupling strengths $g\propto 1/\sqrt{V_{m}}$ and effectively overcome the adversarial effects of Ohmic dissipation and open-cavity radiation losses. By realizing a truly quantum electrodynamic regime of light-matter interaction under ambient conditions and at the nanoscale, the plasmonic platform has opened up a compelling vista of opportunities for fundamental study and technological exploitation alike \cite{xiong2021room}.

In this perspective, we highlight a number of recent studies that, in our opinion, reveal the remarkable potential of nanoplasmonic cQED architectures for driving quantum technologies down to nanoscale spatial and ultrafast temporal regimes, whilst elevating them to ambient temperatures. Our perspective encompasses innovative proposals for ultrasensitive biochemical detection, driving ultrafast single-photon emission and achieving dynamic, multipartite entanglement in the strong coupling regime, with a notable emphasis on the use of industry-grade devices. We conclude with an outlook on a number of promising research directions in quantum nanoplasmonics for the current noisy intermediate-scale quantum (NISQ) era and beyond.

\section{Quantum Plasmonic Immunoassay Sensing}

\begin{figure}
  \centering{\includegraphics[width=0.8\linewidth]{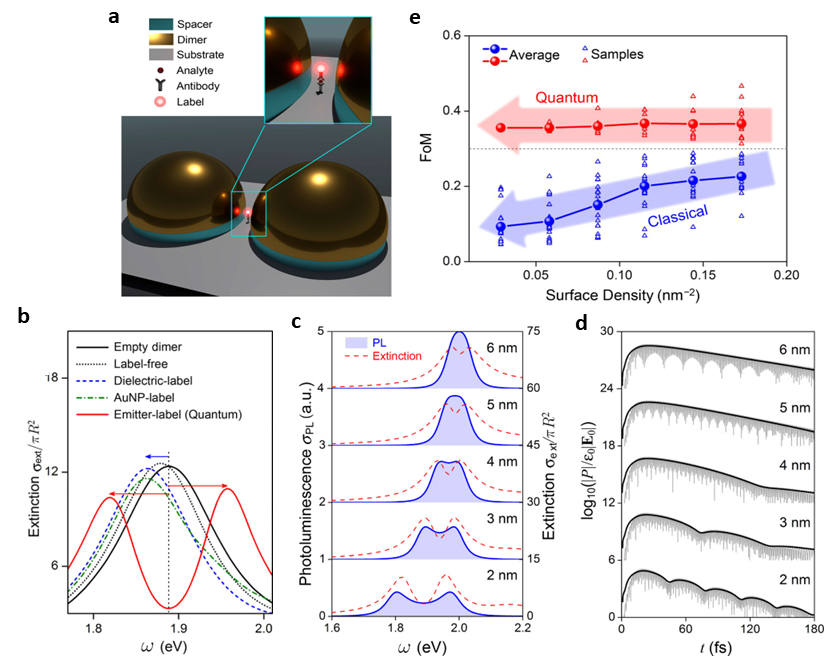}}
  \caption{Quantum plasmonic immunoassay sensing. (a) Schematic diagram of the strong-coupling-enabled sensing device featuring a single analyte-emitter complex at the plasmonic hotspot. Key components are labelled.
  (b) Comparison of the sensing performance of the strong coupling (splitting-type) quantum sensor with classical (shifting-type) plasmonic sensors that feature dielectric and gold nanoparticle (AuNP) labels.
  (c) Photoluminescence (PL, blue solid line) and extinction (red dashed line) spectra for dimers with gap sizes of 2 -- 6 nm and a fixed emitter transition dipole moment of 20 D.
  (d) Time-dependent polarization $P(t)$ of the single emitter label and corresponding envelopes (black solid lines), normalized in accordance with the incident field amplitude $E_{0}$ and plotted in a logarithmic representation.
  (e) Sensing figure of merit (FoM) in classical and quantum regimes as a function of the surface density of analyte-emitter complexes. The FoM is defined by the ratio $\textrm{FoM} = \int \vert \sigma_{\textrm{ext}} - \sigma^{0}_{\textrm{ext}} \vert d\omega / \int \sigma^{0}_{\textrm{ext}}d\omega$, where $\sigma_{\textrm{ext}}$ and $\sigma^{0}_{\textrm{ext}}$ are the extinction cross-sections of the dimer-multianalyte system and free dimer respectively. The triangles indicate the FoM realized in each simulation with a random distribution of complexes, while the spheres indicate the average value for each surface density.
  Adapted with permission from Kongsuwan {\it et al.} \cite{kongsuwan2019quantum}. Copyright 2019, American Chemical Society.}
  \label{fig:Fig1}
\end{figure}

Kongsuwan {\it et al.} \cite{kongsuwan2019quantum} have proposed an innovative, quantum plasmonic immunoassay sensing (QPIS) strategy, which embeds biochemical sensing with recently demonstrated room-temperature strong coupling in nanoplasmonic cavities, enhancing the sensitivity down to the single-molecule limit. Their system design is schematically illustrated in Fig. {\ref{fig:Fig1}}a and involves four main components: (i) a dimer of gold hemispheres acting as a plasmonic resonator, (ii) an antigen as the “analyte” to be detected, (iii) two antibodies that are paired with the target antigen, and (iv) a sensing label that is chemically linked to the antibody–antigen–antibody complex. For a strong-coupling immunoassay scheme, a quantum emitter (such as a quantum dot or dye molecule) is employed as the sensing label, but could, alternatively, be a dielectric or plasmonic nanoparticle, corresponding to a more conventional, purely classical plasmonic immunoassay.

Fig. {\ref{fig:Fig1}}b demonstrates the sensing capability of the quantum plasmonic scheme for a single analyte-emitter complex, via a comparison of the extinction spectrum of the strongly coupled system with those of conventional (classical) plasmonic immunoassays. The spectra of the conventional plasmonic immunoassays exhibit a characteristic shift of the optical resonance. In contrast, the quantum immunoassay shows the characteristic feature of the strong coupling regime, namely a dual-peak structure arising from Rabi splitting. This is, in effect, a bidirectional spectral modification offering a higher sensitivity relative to conventional immunoassays. The extinction spectrum is complemented by an analysis of the photoluminescence lineshape of the emitter labels in Fig. {\ref{fig:Fig1}}c, where for gap sizes below 4 nm, splittings can be observed in both the photoluminescence and extinction spectra, unambiguously confirming strong coupling. Perhaps the most striking manifestation of this quantum regime can be found in the time-dependent polarization $P(t)$ of the emitter, shown in Figure {\ref{fig:Fig1}}d, where ultrafast Rabi oscillations with a period of less than 40 fs are observed, reflecting the coherent energy cycling between plasmons and emitter. Note that although the requisite few-nanometer gaps may preclude access for large biomolecules, recently proposed substrate-engineering principles \cite{xiong2022control,xiong2023substrate} could facilitate elevation of the plasmonic hotspot to the upper end of the gap, thereby rendering it more accessible and making QPIS feasible for large molecules also.   

Of course, most practical scenarios will entail a multitude of emitter-labelled analyte complexes distributed in a serum medium. Fig. {\ref{fig:Fig1}}e presents the results of a statistical study on the extinction spectra for different average surface densities of randomly dispersed analyte-emitter complexes in the vicinity of the hemisphere dimer. Although no clear distinction between the two regimes can be made at large surface densities, the sensing figure of merit degrades rapidly with analyte concentration for classical sensors, yet in the quantum regime remains almost constant, being dominated by the dynamics of only a single emitter located at the plasmonic hotspot.
The results suggest that towards single-analyte detection, the QPIS (splitting-type) protocol unambiguously outperforms more conventional (shifting-type) plasmonic sensors, and establishes an exciting paradigm for further research.

\section{Near-field-driven Single-photon Emission and Multipartite Entanglement}

\begin{figure}
  \centering{\includegraphics[width=1.0\linewidth]{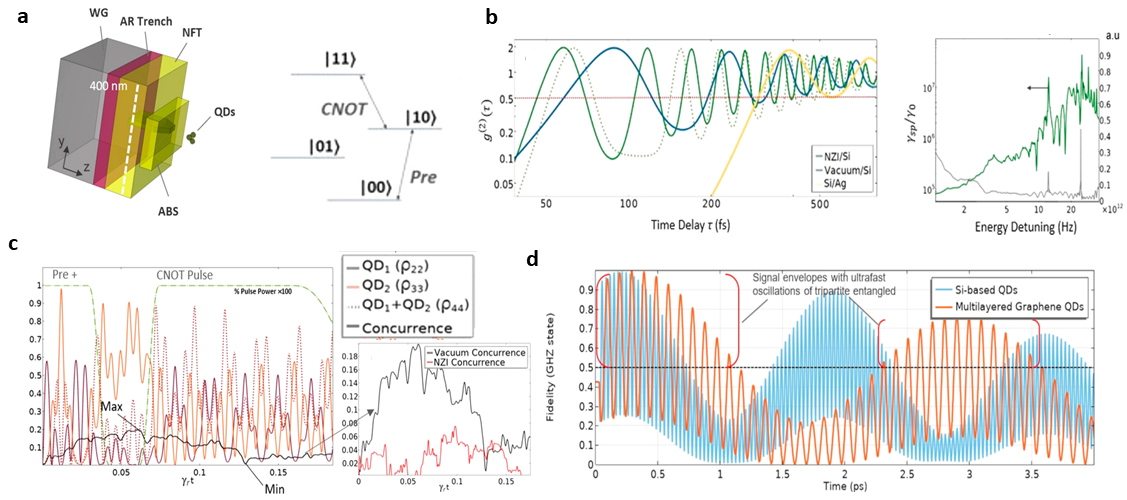}}
  \caption{Single-photon emission and multipartite entanglement of near-field-excited quantum dots.
  (a) Left: Schematic of the near-field transducer (NFT) used to excite Si quantum dots (QDs). A single-mode photonic waveguide (WG) excites an antisymmetric surface plasmon mode in a tapered, metal-insulator-metal device comprising gold and silicon dioxide layers. An antireflective (AR) trench is used to improve optical efficiency. The QDs are located at some fixed depth relative to the air-bearing surface (ABS) of their host medium. Right: Energy levels of a bipartite system comprising two QDs, each modelled as a two-level system. The bipartite ground state is $\vert 00\rangle$, the states $\vert 01\rangle$ and $\vert 10\rangle$ each feature a single excited QD, whereas in the state $\vert 11\rangle$, both QDs are excited.  
  (b) Left: second-order intensity autocorrelation function for three different materials embedding a single QD, including a near-zero index (NZI) film and vacuum over silicon (Si) substrates, as well as a Si film on silver (Ag). Values below 0.5 indicate that single-photon emission dominates. The grey dotted curve corresponds to an order of magnitude reduction in the NFT input power, thereby altering the rate of single-photon emission. Right: Enhancement in the QD spontaneous emission rate $\gamma_{sp}$ relative to its free-space value $\gamma_{0}$, as a function of excited-state detuning.
  (c) Left: Time-dependent probabilities of populating the individual QD excited states (in vacuum), along with the concurrence, in the presence of a sequence of pulses from the NFT. Note that $\gamma_{r} = 8\times 10^{11}$ s$^{-1}$. Right: Concurrence for QDs in vacuum and NZI media.
  (d) Fidelity with which the Greenberger-Horne-Zeilinger state is produced in a tripartite system of Si QDs in both Si and graphene material environments. Values above 0.5 certify genuine tripartite entanglement.
  (b,c) Adapted with permission from Bello {\it et al.} \cite{bello2020cavityfree}. Copyright 2020, American Chemical Society.
  (a,d) Adapted with permission from Bello {\it et al.} \cite{bello2022near}. Copyright 2022, American Chemical Society.}
  \label{fig:Fig2}
\end{figure}

Recently, Bello {\it et al.} \cite{bello2020cavityfree,bello2022near} have theoretically explored a scheme for enabling single-photon emission and multipartite entanglement in Si quantum dots (QDs), harnessing the sub-diffractive focusing of optical fields by a commercial, plasmonic near-field transducer (NFT) device schematized in Fig. {\ref{fig:Fig2}}a. Originally explored for heat-assisted magnetic recording applications in the hard-drive industry, the latest developments in NFT technology allow full integration with photonic waveguides and have made possible the nanofocusing of light to regions on a scale of 50 nm $\times$ 50 nm, with an optical energy efficiency as high as $15\%$. As evidenced by the theoretical, intensity autocorrelation function in Fig. {\ref{fig:Fig2}}b, the enhanced light-matter interaction gives rise to single-photon emission at a rate of 10 -- 100 fs$^{-1}$, manipulable via modulation of the NFT input power and selection of the embedding medium for the QDs. Particularly high coupling strengths were identified for dots embedded in a near-zero-index film, leading to an increased rate of Rabi oscillations between the QD ground and excited states, as well as Purcell-enhanced spontaneous emission rates by a factor as large as $10^{7}$ relative to free space.

Crucially, the transducer near-field can also be harnessed to generate dynamic bipartite entanglement, enabling one of the simplest two-qubit quantum logic gates, namely the controlled NOT (CNOT) gate. The degree of entanglement between the two qubits, whose energy levels are shown in Fig. {\ref{fig:Fig2}}a, can be systematically characterized via the concurrence $C$, with a maximum value of $C=1$ realized when, for instance, the system can be prepared in the Bell states $(\vert 01\rangle \pm \vert 10\rangle)/\sqrt{2}$ or
$(\vert 00\rangle \pm \vert 11\rangle)/\sqrt{2}$ \cite{horodecki2009quantum}.
Fig. {\ref{fig:Fig2}}c shows the concurrence for two near-field-excited QDs in vacuum. Here, a single pulse from the NFT induces Rabi oscillations between the two-qubit ground state $\vert 00\rangle$ and the state $\vert 10\rangle$ featuring a single-qubit excitation. A second pulse from the NFT populates the state $\vert 11 \rangle$, after switching off the prepulse. Although $C$ remains much smaller than 1, it is similar to many proof-of-concept values reported using plasmonic nanocavities \cite{xiong2021room}, and extends previous studies of plasmon-mediated entanglement into the strong coupling regime. Moreover, the achieved values of concurrence could potentially be improved via optimized NFT pulsing schemes and QD positioning in the near-field environment, or resonator designs based on alternative, low-loss plasmonic materials. Of note, the enhanced light-matter interaction facilitated by a near-zero-index host medium does not translate to improved entanglement between the qubits in this particular case, where the maximum achievable concurrence (about 0.07) is much smaller than for vacuum (about 0.2), in turn manifesting the need for an improved control over the coherences among the bipartite states.

This work has been further extended to the generation and dynamic modulation of three-qubit entanglement under ambient conditions \cite{bello2022near}. As demonstrated for QDs embedded in silicon and multilayer graphene in Fig. {\ref{fig:Fig2}}d, repeated excitation of a tripartite Greenberger-Horne-Zeilinger state, $(\vert 000\rangle + \vert 111\rangle)/\sqrt{2}$, can be achieved with fidelities exceeding 0.5, and even approaching the maximum value of 1.0 at certain instants of time. Changes in the position of the QDs, or the choice of host medium and its thickness, may also alter the rate of oscillations and prove beneficial for optimizing a particular design.
The ability to adeptly manipulate quantum superpositions and entanglement on nanometric spatial scales and ultrafast timescales will inevitably prove instrumental to the implementation of quantum networking concepts and error-correcting protocols for fault-tolerant QIP systems in chip-scale environments.

\section{Outlook}

In this perspective, we have focused on recent progress in harnessing nanoplasmonic architectures for strong-coupling-enabled sensing, ultrafast single-photon emission and multi-qubit quantum logic under ambient conditions. Spurred by its promise, the field of plasmonic quantum nanotechnologies remains dynamic and ever-evolving, driven by its bespoke confluence of nanoscale compactness and integration potential, ultrafast functionality as well as room-temperature viability.

The inherently small physical size of quantum plasmonic devices is itself an important quality. Their nanoscale compactness is naturally compatible with the drive for miniaturization in photonics and offers prospects for the development of chip-scale technologies that densely integrate many interconnected, quantum-optical components. This potential has been further enhanced by two complementary developments. Firstly, the emergence of plasmonic nanocircuitry \cite{wu2017onchip,ochs2021nanoscale}, which paves the way towards fully integrated, electrically-operable platforms for nanoplasmonic cavity and waveguide quantum electrodynamics. Secondly, the recent proposal of dielectric engineering strategies for plasmonic hotspots and the control of plasmon-exciton strong coupling in low-dimensional semiconductors \cite{xiong2022control,xiong2023substrate}, which is fully compatible with well-established, top-down lithographic fabrication approaches.

The plasmon-exciton dynamics facilitated by plasmonic nanocavities evolves on remarkably short timescales of merely tens of femtoseconds (generally limited by the cavity plasmon lifetime itself). As we have seen, such ultrafast dynamics out-competes decoherence effects and facilitates the coherent manipulation of quantum emitters at room temperature \cite{bello2020cavityfree,bello2022near}. Another issue that merits exploration, however, surrounds the potential of plasmon-exciton strong coupling as a basis for time-resolved spectroscopy. Recent works have already seeded a notion of ``strong coupling spectroscopy'' \cite{gross2018near,park2019tip}, where single-emitter excitations are manifested and identified via the spectral signatures of their strong coupling with cavity photonic and plasmonic modes. Yet, the temporal dynamics of a strongly-coupled system should also encode information pertaining to the plasmon and matter excitations, and may thereby establish a novel means for spectroscopic characterization of electronic and phononic excitations. Notably, a promising synergy is emerging between nanoplasmonics and attosecond physics \cite{dombi2020strong}, which may deliver and advance on this potential in the near-future.

Plasmonic cQED schemes have traditionally relied on stationary cavity structures, etched into bulk media or requiring careful placement of emitters using external means \cite{baranov2018novel,baumbergetal2019rev}. In contrast, industry-grade NFTs are non-integrated devices, allowing rasterization across arrays of quantum emitters lying tens of nanometers away from their throughput end, as opposed to the extreme resonator-emitter proximity demanded by current plasmonic cQED experiments. This eases fabrication constraints while not sacrificing the ability to efficiently control picosecond-scale dynamic entanglement of qubits. Ultimately, NFT technology may improve the compactness, accessibility and scalability of QIP devices in the NISQ era, opening a route to quantum plasmonic memory schemes and driving quantum nanotechnologies ever closer to the commercial realm.

\backmatter

\section*{List of Abbreviations}

CNOT, controlled NOT; cQED, cavity quantum electrodynamics; NFT, near-field transducer; NISQ, noisy intermediate-scale quantum; QD, quantum dot; QIP, quantum information processing; QPIS, quantum plasmonic immunoassay sensing.

\section*{Declarations}

\subsection*{Availability of Data and Materials}

Data sharing is not applicable to this article as no datasets were generated or analysed during the current study.

\subsection*{Competing Interests}

The authors declare that they have no competing interests.

\subsection*{Funding}

DDAC and OH gratefully acknowledge funding from Science Foundation Ireland via Grant No. 18/RP/6236.

\subsection*{Authors' Contributions}

Both authors contributed to the preparation of the manuscript and approved the final version.

\subsection*{Acknowledgements}

Not applicable.

\bibliography{bibliography}

\end{document}